# Dual-energy CT Reconstruction from Dual Quarter Scans

Wenkun Zhang, Ningning Liang, Linyuan Wang, Ailong Cai, Zhizhong Zheng, Chao Tang, Yizhong Wang, Lei Li*, Bin Yan, Guoen Hu

***Abstract*—Compared with conventional single-energy computed tomography (CT), dual-energy CT (DECT) provides better material differentiation but most DECT imaging systems require dual full-angle projection data at different X-ray spectra. Relaxing the requirement of data acquisition is a particularly attractive research to promote the applications of DECT in a wide range of imaging areas. In this work, we design a novel DECT imaging scheme with dual quarter scans and propose an efficient method to reconstruct the desired DECT images from dual limited-angle projection data, which enables DECT on imaging configurations with half-scan and largely reduces scanning angles and radiation doses. We first study the characteristics of image artifacts under dual quarter scans scheme, and find that the directional limited-angle artifacts of DECT images are complementarily distributed in image domain because the corresponding X-rays of high- and low-energy scans are orthogonal. Inspired by this finding, a fusion CT image is generated by integrating the limited-angle DECT images of dual quarter scans. This strategy largely reduces the limited-angle artifacts and preserves the image edges and inner structures. Utilizing the capability of neural network in the modeling of nonlinear problem, a novel Anchor network with single-entry double-out architecture is designed in this work to yield the desired DECT images from the generated fusion CT image. Experimental results on the simulated and real data verify the effectiveness of the proposed method.**

## I. Introduction

DUAL-energy computed tomography (DECT) is widely used in advanced medical imaging [1], [2], security inspection [3], [4], and nondestructive testing [5]. Compared with single-energy CT, DECT provides two sets of attenuation measurements at two different energy spectra and enables enhanced material characterization by exploring the interdependence of X-ray attenuation and photon energy [6]. Most DECT imaging systems require two full-scan projection datasets [7]. As such, the projection data at each energy spectrum should be collected over 360° as a full-scan configuration, limiting its applications to wide imaging areas that cannot meet this requirement. In this study, we design a novel flexible DECT imaging scheme and propose an efficient reconstruction method to relax the requirement of data collection and reduce the scanning angles and radiation doses for DECT.

The reconstruction methods for DECT can be categorized into three types, namely, direct reconstruction [8], projection domain- [9], and image domain-based [10] methods. The direct reconstruction method can model the imaging procedure and directly reconstruct the basis materials from the projection data but is computationally expensive due to the repeated forward projection and backprojection [11]. The projection domain-based method first obtains the sinograms of basis materials from dual-energy projections and then reconstructs the basis materials. This method can correct the beam-hardening artifacts but requires accurate system calibration to obtain dual-energy measurement data on the same projection ray, which is a challenge for CT imaging systems, such as dual-source and fast kVp-switching configurations [12]. Compared with the two former methods, the image domain-based method decomposes basis materials from the readily available reconstructed dual-energy CT images and is more convenient to embed into the imaging system [13], [14]. Thus, this method has become the most widely used method in commercial CT scanners. For the image domain-based method, the reconstructed images with promising quality are necessary for the following material decomposition [15]. The missing scanning angles/views in either energy may degrade the reconstructed CT images, subsequently destroying material decomposition. Thus, most existing research studies are carried out on the basis of the DECT images reconstructed from the full-scan projection data to obtain high-quality results [16]–[18].

Recently, scanning schemes with measurements less than full angles/views begin to be investigated by researchers. Shen et al. divided the circular trajectory of CT into several arcs with different tube voltages and proposed a segmental multienergy CT reconstruction method [19]. Wang et al. proposed a reconstruction approach with one full scan and a second sparse-view scan by introducing a similarity matrix between the high- and low-energy images [20]. On the basis of this method, Petrongolo et al. further designed a primary modulation DECT scheme and reconstructed images from dual-energy sparse-view projections [21]. The above methods have introduced beneficial results for DECT but still require a circle of full-angle scan and become disabled in the imaging scheme with scanning angle less than 360°. To solve this problem, Xing et al. first simulated a DECT less than full-

This work is supported by the National Natural Science Foundation of China (No. 61601518) and the National Science Foundation for Post-doctoral Scientists of China (No. 2019M663996). (Corresponding author: Lei Li)

The authors are with Key Laboratory of Imaging and Intelligent Processing of Henan Province, PLA Strategic Support Force Information Engineering University, Zhengzhou, 450002, P.R.China. (E-mail: zhangwenkun1991@163.com; ning_ning216@163.com; wanglinyuanwly@163.com; cai.ailong@163.com; zhengzoom@126.com; tangchao10@163.com; wyzwyz0101@163.com; leehotline@163.com; ybspace@hotmail.com; guoenxx@163.com)

angle scan and performed experiments on the digital phantom [22]. Chen et al. then proposed an optimization-based one-step method to enable DECT on a short-scan configuration [23], which provides a partial-scan solution for DECT but belongs to the one-step method with high computation complexity. Zhang et al. further proposed an image domain-based reconstruction method with a half-scan plus a second limited-scan [24]. Although this method obtained promising results under the partial-scan scheme, its second scan arc over 180° must be large enough to generate acceptable limited-angle results. In the present study, we aim to enable DECT imaging within a half-scan range to promote the partial-scan scheme in DECT. As shown in Fig. 1, the scanning scheme consists of dual quarter scans. The X-ray tube is tuned to change the tube voltage at the angle of 90°, and the measurement data of dual quarter scans with high- and low-energy spectra can be obtained. The designed scanning scheme largely reduces the requirement of scanning angles and measurement data, and provides a flexible DECT scanning scheme for a wide range of imaging configurations with partial-scan, such as C-arm type systems. However, for this scanning scheme, two arcs generate two serious limited-angle problems with 90° scanning angle. The limited-angle problem is an intractable issue in practice as we know and has not been solved well even in conventional CT [25]. Thus, in this work, obtaining promising reconstruction CT images from dual limited-angle measurement data becomes a challenging issue.

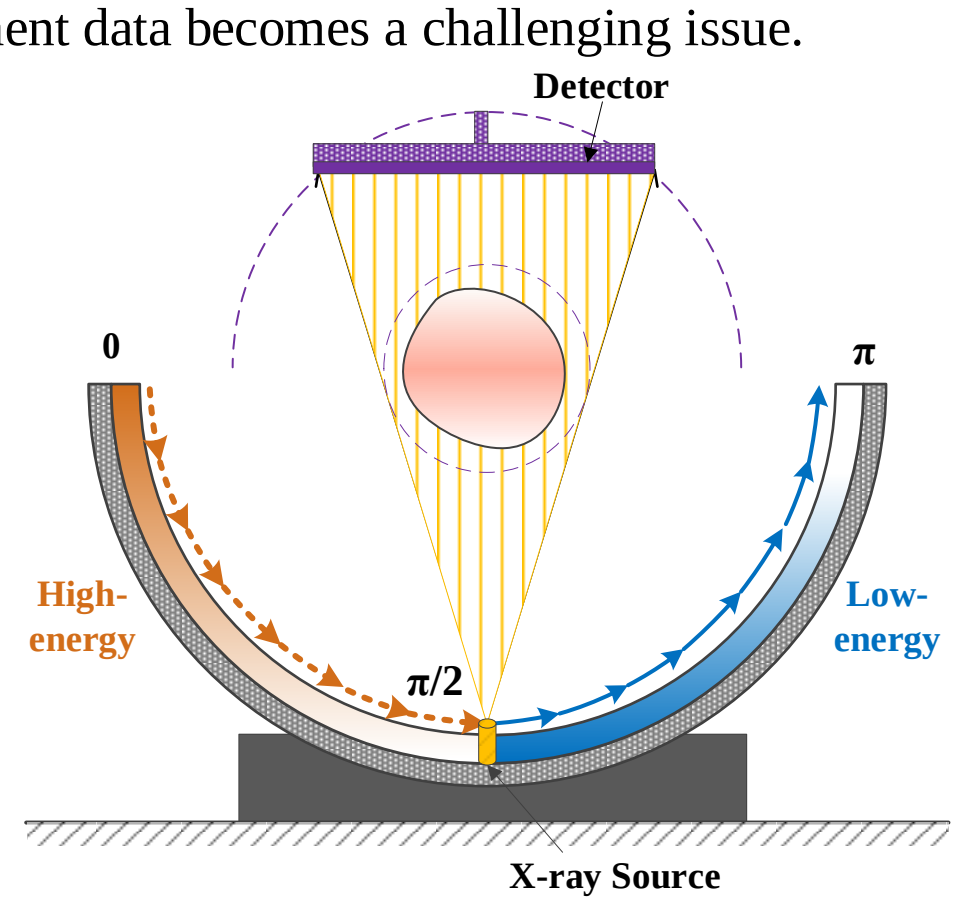

Fig. 1. DECT imaging scheme with dual quarter scans.

For conventional single-energy CT, iterative methods with regularization terms have been studied to suppress the limited-angle artifact by exploring additional prior/empirical knowledge, such as total variation (TV) [26], [27], wavelet tight frame [28], nonlocal block similarity [29], low rank [30], and dictionary learning [31]. These methods obtain beneficial results from limited-angle measurement data, but when the scanning range is very narrow, their reconstruction quality largely declines. Theoretical research has also been done to analyze the characteristic of limited-angle artifacts. Quinto discussed the feature of limited-angle problem and illustrated that the details not tangent to the X-ray lines are difficult to reconstruct [32]. Frikel et al. verified that the limited-angle artifacts may appear only along the lines that are tangent to the singularities of the original object via microlocal analysis [33]. The strength of limited-angle artifacts was also characterized mathematically [34]. These theoretical studies provide us useful guidance in reducing limited-angle artifacts on the basis of the characteristics of limited-angle artifact. However, maximizing the role of these studies still needs further research. Deep learning methods recently receive increasing attention in CT imaging [35]–[37]. Zhang et al. firstly utilized convolutional neural network to extract and reduce limited-angle artifacts [38]. Gu et al. proposed a multiscale wavelet domain residual learning to reduce artifacts [39]. Zhao et al. used generative adversarial network to realize sinogram inpainting for limited-angle reconstruction [40]. The learning-based methods receive promising results in conventional CT but heavily rely on the big training data and ignore the theoretical characteristic analysis for the limited-angle artifacts. Nevertheless, we still believe that deep learning will play an important role in the solution of nonlinear problems those have not been depicted by mathematical models or have been modeled by mathematics but can further optimize their accuracy and speed. Back to reconstruction problems of this work, few deep learning-based strategies are made to simultaneously solve the dual limited-angle problems.

In this study, we will propose an efficient method for the designed DECT scheme with dual quarter scans by simultaneously exploring the characteristics of limited-angle artifacts and taking the advantage of neural network in nonlinear mapping. On the basis of the theoretical analysis of artifact characteristic, we first illustrate that the limited-angle artifacts of high- and low-energy reconstruction images are complementarily distributed in image domain because the designed scanning scheme in our work presents an orthogonal geometry for the corresponding X-ray lines of dual quarter scans. Inspired by this finding, a fusion CT image is generated by integrating the dual limited-angle reconstruction images. The fusion CT image will largely reduce the limited-angle artifacts caused by the missing of projection data and show a good structure and edge consistency with the ground truth image. A novel neural network with single-entry double-out architecture is specifically designed in this work to finally yield the desired DECT images from the generated fusion CT image. The experimental results verify the effectiveness of the proposed method.

## II. Methodology

### A. *Limited-angle Artifact Characteristic*

The measurement data of limited-angle CT are restricted to lines in a limited-angle range (less than 180°). As the projection data are incomplete, the limited-angle problem is highly ill-posed. Standard CT reconstruction methods will not obtain a reliable solution and the reconstruction image usually suffers from serious artifacts. Filtered-backprojection (FBP) method [41] is the most widely used method in practice. Considering the well-known FBP reconstruction from limited-angle measurement data, some remarkable phenomena can be observed in CT image.

As shown in Fig. 2, a homogeneous disk phantom (Fig. 2(a)) is scanned within a limited-angle range of $[-\theta/2, \theta/2]$. Fig.

2(b) shows the reconstruction image obtained via FBP method without enforcing the negative pixel values to be zero. Fig. 2(c) provides the geometry map of reconstruction result for the characteristic analysis of image artifact. We can find that the limited-angle artifacts show a related directional property for a directional angular coverage in scanning. First, the direction of streak artifact is consistent with the end of the limited-angle scanning range (denoted by the gray dashed lines in Fig. 2(c)), which indicates that the direction of streak artifact changes with the scanning angles. Second, additional black and white artifacts are created in the region where the X-ray lines are tangent to the original object, as shown by the black and gray regions in Fig. 2(c). Notably, the additional black artifacts with negative gray values are symmetrically distributed with the additional white artifacts with positive gray values. Third, the image edge of reconstruction image is maintained well in the vertical direction of scanning X-rays, as shown by the red line in Fig. 2(c), but not in the parallel direction. In our work, we aim to solve the dual limited-angle problems for the designed dual quarter scans scheme by utilizing the above specific characteristics of limited-angle artifacts.

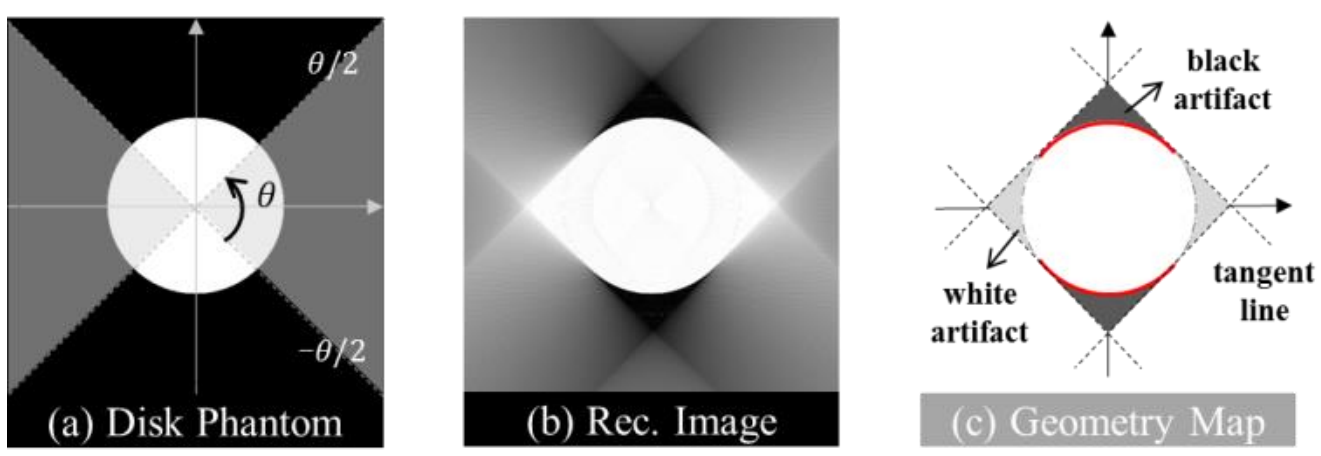

Fig. 2. Limited-angle reconstruction: (a) scanning phantom, (b) reconstructed image via FBP method without enforcing the negative pixel values to be zero, (c) geometry feature map of limited-angle artifact. The display window of (a) and (b) is [0, 0.1] and [−0.1, 0.1], respectively.

### *B. Fusion CT Image Formation*

For the designed DECT imaging scheme in our work, the first quarter scan at high tube voltage collects measurement data from 0 to $\pi/2$, whereas the second quarter scan at low tube voltage collects measurement data from $\pi/2$ to $\pi$. The corresponding X-ray lines of dual quarter scans are $\pi/2$ apart, i.e., they are orthometric to each other.

As shown in Fig. 3, the disk phantom assumed to be composed of tissue material is scanned under the proposed DECT scheme with 80 and 140 kVp tube voltages (Figs. 3(a) and 3(b)). Dual-energy sinograms are collected under dual quarter scans (Figs. 3(c) and 3(d)). FBP method is applied to reconstruct DECT images (Figs. 3(e) and 3(f)) from dual limited-angle projections. (Note that the FBP reconstruction results used in the procedure of generating fusion CT image do not enforce the negative pixel values to be zero, which applies all the same instances in this work.) Based on the analysis of Sec. IIA, the DECT reconstruction results show that the dual limited-angle problems of the designed scanning scheme have the following typical characteristics.

1) The limited-angle artifacts of dual quarter scans are distributed along the almost same scanning boundaries in circular trajectory (denoted by white dashed lines).

2) The position of black artifacts with negative values in the reconstruction image of first quarter scan corresponds to the position of white artifacts with positive values in the reconstruction image of second quarter scan (denoted by gray arrows), and the same is observed for the white artifacts under first quarter scan and the black artifacts under second quarter scan (denoted by white arrows).

3) The distorted structures in the first quarter scan remains good in the second quarter scan (denoted by yellow arrows), whereas the distorted structure in the second quarter scan remains good in the first quarter scan (denoted by blue arrows).

Inspired by the above typical analysis, we utilize these complementary characteristics of dual quarter scans on their reconstruction images to reduce the limited-angle artifacts. A fusion CT image is generated in this work by integrating the limited-angle high- and low-energy CT images. Let $x_H$ and $x_L$ represents the limited-angle high- and low-energy CT image, respectively. The fusion CT image $x_F$ can be calculated using the equation:

$$x_F = \alpha x_H + (1-\alpha) x_L, \tag{1}$$

where $\alpha$ is the parameter to balance the gray value of the image pixel at the same position of high- and low-energy CT images. Fig. 3(g) represents the generated fusion CT image with $\alpha = 0.5$ based on the limited-angle high- and low-energy CT images. Fig. 3(h) denotes the bias map of fusion CT image with the reference image. The reference image is formed by combining the DECT images reconstructed from full-angle projections using the same generation method of fusion CT image. We can find that the formation of fusion CT image largely reduces the streak artifacts caused by the missing of scanning angles, and efficiently suppresses the additional black and white artifacts around the image object. The object edges of fusion CT image are also restored with high quality and become clearer than those of the initial limited-angle DECT images (denoted by red arrows). In theory analysis, the integration of the limited-angle DECT images corresponds to that of the limited-angle dual-energy sinograms because the image reconstruction of DECT is a linear operation. The limited-angle artifacts are caused by the singularities of truncated sinogram and locally distributed in image-domain. The combined sinogram becomes smooth and continuous and are only truncated at the junction of dual-energy sinograms. In this way, the singularities of combined sinogram are largely reduced, resulting in the reduction of image artifacts caused by the deficiency of projection data [42]. Anyway, the formation of fusion CT image enables the limited-angle problems of dual quarter scans expect to be solved in a concise strategy, which is a novel observation for the DECT reconstruction of dual quarter scans scheme in this work.

We can also find from Figs. 3(g) and 3(h) that the fusion CT image still contains little artifacts and gray value shift, but these do not affect the fusion CT image considered as a pilot image with promising quality to guide the generation of high- and low-energy CT images. Considering that no mathematical model is established to yield DECT images from fusion CT image, we design a novel neural network in this work to learn their mapping model for this specific imaging problem by utilizing the capability of neural network in the solution of nonlinear problem that has not been modeled by mathematics.

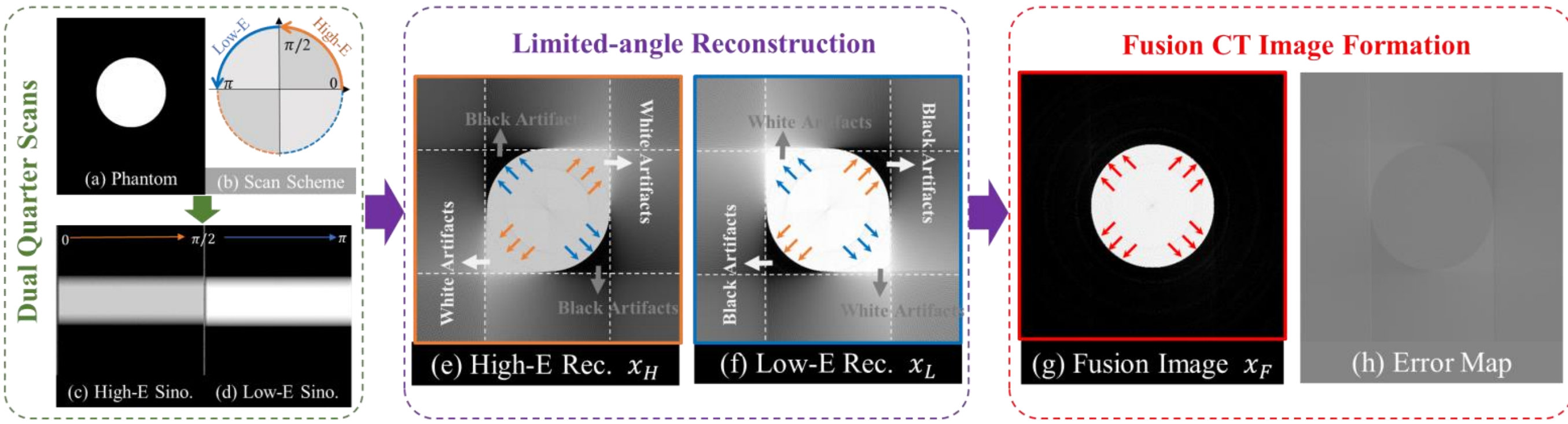

Fig. 3. Formation of the fusion CT image for the proposed dual quarter scans scheme: (a) scanning phantom composed of tissue material and (b) dual quarter scans scheme. (c) High- and (d) low-energy sinograms. (e) High- and (f) low-energy reconstruction results with display window of [−0.02, 0.02]. (g) Fusion CT image with display window of [0, 0.02] generated by integrating limited-angle DECT images. (h) Bias map of fusion CT image with the reference image. The reference image is formed by combining the DECT images reconstructed from full-angle projections using the same generation method of fusion CT image. The display window of (h) is [−0.02, 0.02].

### *C. Anchor Network (AnNet)*

The network of the proposed method is designed as a single-entry double-out architecture on the basis of the architecture of Unet convolutional network [43]. The designed network consists of a contracting path and two expansive paths. Two expansive paths point to the desired high- and low-energy CT images. The designed network is illustrated in Fig. 4. Each blue box represents a multichannel feature map. The image size and the number of feature channel is denoted by normal and bold numbers, respectively, beside the data box. The gray box represents the copied feature maps. The green arrows represent the convolution followed by a rectified linear unit (ReLU). The red arrows represent the max pooling, and the yellow arrows denote the upsampling convolution. The purple arrows represent the 1 × 1 convolution, and the gray line represents the skip connection. The new network is named as Anchor network (AnNet) in our work in accordance with its single-entry double-out architecture.

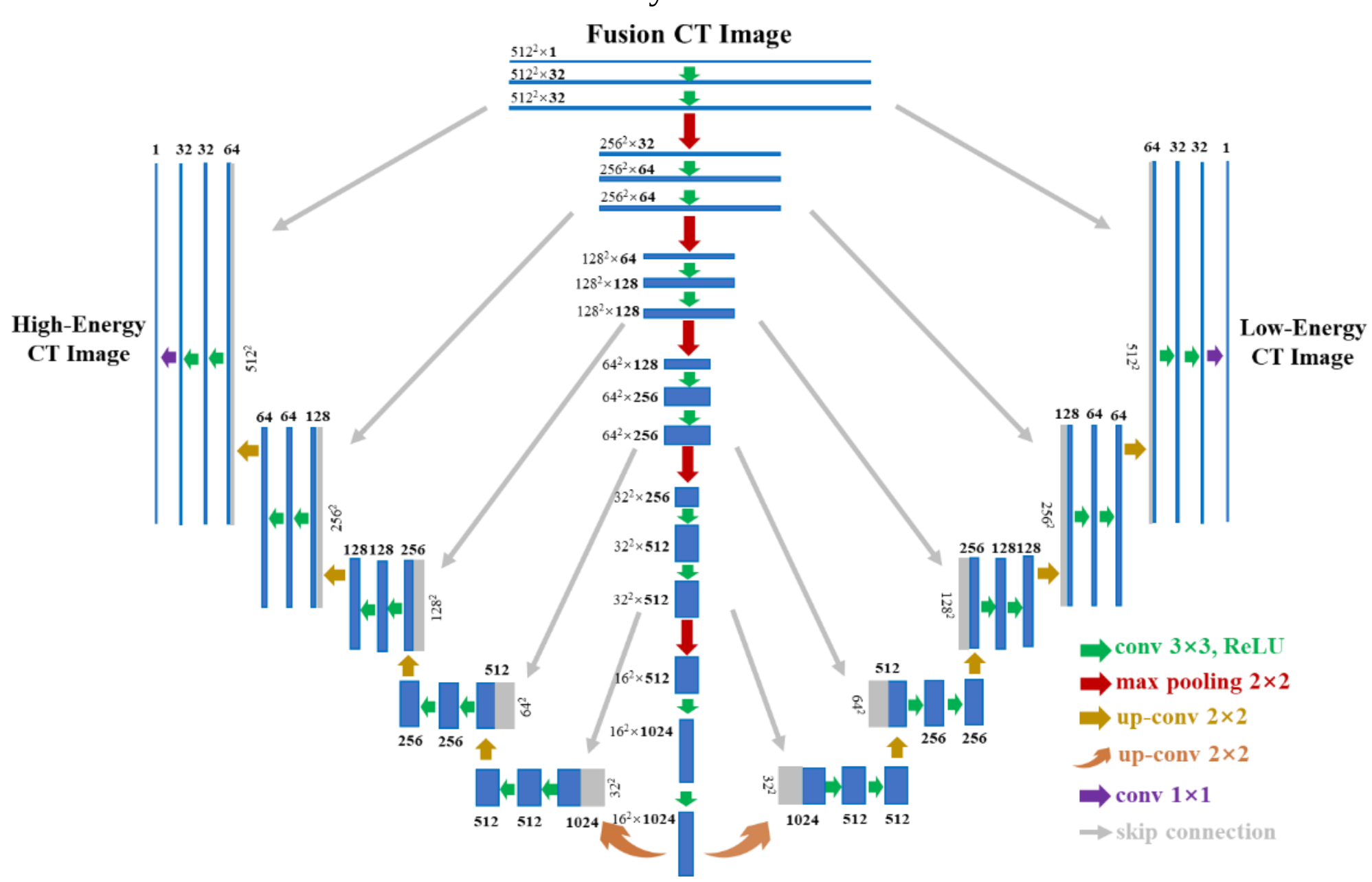

Fig. 4. Anchor network (AnNet). Each blue box represents a multichannel feature map. The image size and the number of feature channels is denoted by normal and bold numbers, respectively, beside the data box. The gray box represents the copied feature maps. The arrows in different colors denote the different operations of the network.

In this network, the contracting path (middle path) consists of six modules. In each module, the repeated operation of two 3 × 3 convolutions followed by a ReLU is applied on the input data. Then, a 2 × 2 max pooling with stride 2 is used for the downsampling operation to extract abstract features except the bottom module of contracting path. The max pooling doubles the number of feature channels. Two expansive paths (right and left paths) are the same except their output images. Each expansive path consists of five modules. In each module, an upsampling of feature map and a 2 × 2 convolution are applied on the output data of the last module, which halves the number of feature map. A skip connection with feature map is introduced to provide high-resolution detailed information from the corresponding module in the contracting path. The repeated application of two 3 × 3 convolutions followed by a ReLU is then applied on the concatenated data. At the final layer of each expansive path, a 1 × 1 convolution is used to map the 32 channels of feature vector to the desired CT image at high- or low- energy spectra.

The network loss consists of high- and low-energy losses. The training dataset $D = \{(x_1, y_1^L, y_1^H), \ldots, (x_N, y_N^L, y_N^H)\}$, where $x_i$, $y_i^L$, and $y_i^H$ represents the input fusion CT image, the desired low- and high-energy CT image, respectively, and $i$ and $N$ denotes the index and total number of training samples, respectively. The parameter set of AnNet is represented by $\Phi$. Then the cost function in terms of the mean squared error is defined as:

$$\begin{aligned} J(D;\Phi) &= \frac{1}{N}\sum_{i=1}^{N}\{\mathcal{L}(\hat{y}_i^L, y_i^L) + \mathcal{L}(\hat{y}_i^H, y_i^H)\} \\ &= \frac{1}{N}\sum_{i=1}^{N}\{\|\hat{y}_i^L - y_i^L\|^2 + \|\hat{y}_i^H - y_i^H\|^2\}, \end{aligned} \quad (2)$$

where $\hat{y}^{L,H}$ denotes the estimated results of the designed network for the low- and high-energy spectra.

### D. Training Dataset Preparation

The training of AnNet requires the fusion CT and DECT images for the same object. DECT reconstructs images from the polychromatic projections, which are generated via the nonlinear polychromatic model based on the basis material images [46]. Thus, we first collect 1430 pairs of head basis material images (bone and tissue) from 7 patients with the assistance of radiologists from local hospitals and then generate polychromatic projections under different scanning ranges. The mass attenuation coefficients of the basis materials are obtained from the National Institute of Standards and Technology database. Polychromatic spectra of 80 and 140 kVp are generated using the SpekCalc software with an energy sampling interval of 1 keV. The source-to-object and source-to-detector distance is 1000 and 1500 mm, respectively. Dual-energy projections are uniformly sampled in 720 views over a circle rotation. Projections are collected using a linear detector that consists of 1024 bins with each pixel of 0.2042 mm. The limited-angle DECT images are first reconstructed using FBP method on the basis of the low- and high-energy projections at $[0, \pi/2)$ and $[\pi/2, \pi)$, respectively. Then, we linearly combine the limited-angle CT images at low- and high-energy spectra on the basis of Eq. (1) to generate the fusion CT image for the network input. The size of reconstructed CT image is $512 \times 512$, and each pixel is $0.2723 \times 0.2723$ mm$^2$. FBP method is also used to reconstruct DECT images from full-angle projections without added noises, which are considered as the network labels. In this way, we can generate 1430 groups of samples on the basis of the collected basis material images, where 1393 and 37 samples are used as training and validation data, respectively, for the simulated data experiment.

Although our effort in generating training data by using the digital phantom can simulate the data generation process of CT scanning, completely narrowing the difference of real and simulated data in statistical distribution is challenging regarding the tube physics, detector physics, and electronics of a practical CT scanner [47]. Therefore, based on the usual strategy of existing learning-based methods [47]–[49], acquiring real data from physical scanners is important to tune the network parameters for the trained network to produce the desired reconstruction results for a specific CT scanner. In this work, the anthropomorphic head phantom is scanned under the tube voltages of 80 and 140 kVp by using a physical CT scanner in our laboratory. DECT images reconstructed from full-angle are considered as the network labels, and the limited-angle DECT images reconstructed from $[0, \pi/2)$ and $[\pi/2, \pi)$ projections are used to generate the fusion CT image for the network input. Among the anthropomorphic head phantom data, 672 and 40 slices of physical phantom are randomly selected as the training and validation data, respectively, to train the AnNet further for real data experiment.

### E. DECT Reconstruction from Dual Quarter Scans

In summary, the proposed DECT method for the designed dual quarter scans scheme consists of two parts: 1) the formation of fusion CT image and 2) the mapping from fusion CT image to DECT images via the AnNet. The first part aims to suppress the limited-angle artifacts and restore the distorted image edges by analyzing the artifact characteristics. The second part is a nonlinear mapping operation via the AnNet, whose main objective is to yield DECT images from the fusion CT image and suppress image noises and residual artifacts. In general, the proposed DECT reconstruction method for the designed dual quarter scans scheme in this work can be summarized as follows.

(1) Reconstruct the limited-angle CT images at high- and low-energy spectra via FBP method on the basis of the limited-angle projections under dual quarter scans.

(2) Generate the fusion CT image by integrating the limited-angle DECT images via Eq. (1).

(3) Calculate the desired DECT images by using the trained AnNet with the fusion CT image as the input data.

## III. Evaluation

The proposed method is compared with the conventional FBP and TV-based methods [50], which are the most widely used analytic and iterative methods in CT imaging. Model solving stops for the iterative method when its iteration reaches 100. We also implement a learning-based method that does not contain the guidance of fusion CT image but directly takes the limited-angle CT image as input data to evaluate the role of fusion CT image in the reconstruction of dual limited-angle problems. In the compared learning-based method, two Unets are independently trained to obtain the desired DECT images. They consider the quarter scan CT image as input image and the full-angle CT image under their corresponding energy as label image. Each Unet consists of a contracting path and an expansive path, and their parameters and layer numbers are consistent with those of the AnNet. The training samples of the learning-based method used are the same with those of the proposed method except their input image.

The simulated data of digital phantom and the real data of physical anthropomorphic head phantom are tested in the experiments with $512 \times 512$ reconstruction image pixels to validate and evaluate the performance of the proposed method under the dual quarter scans scheme. The scanning parameters of the simulated testing data are consistent with those of training data, and Poisson noises with $1 \times 10^5$ incident X-ray photons are added on the projections to simulate image noises.

For the real testing data, one slice of limited-angle projections at high- and low-energy spectra is extracted to perform DECT reconstruction from dual quarter scans. In this work, the reconstruction images of real data are shown at an enlarged field of view with 410 × 410 square pixels. The testing data of the digital and physical phantoms are not part of their training data of the neural network. The slice data around the testing data are also removed from the training dataset. The training iterations of AnNet and Unet are fixed to 300 thousands for the simulated data and 100 thousands for the real data. Parameters of the two networks are initialized using the normal distribution method [44] and updated by the adaptive moment (Adam) algorithm [45], wherein the learning rate is $1 \times 10^{-6}$ and the exponential decay rates for the first and second moment estimates is 0.9 and 0.99, respectively. All computations for learning methods are performed on a workstation with two Intel Xeon E5-2640 v4 CPU 2.4 GHz and four GeForce GTX 1080 Ti GPUs.

For the evaluation of the simulated and real data, the reconstruction results of FBP method based on the full-angle noise-free projection data are considered as the reference images. The parameter $\alpha$ of Eq. (1) is fixed to 0.5 to generate the fusion CT images for simulated and real data. The fusion CT image is shown in this work to reveal its specify effectiveness in artifact suppression and edge preservation. The reconstructed results of different methods based on dual limited-angle projections are compared. A region of interest (ROI) that contains complex structure is magnified for detailed comparison. Bias maps of the reconstruction results are provided in this work to evaluate their differences with the reference images. The line profiles along the direction of limited-angle artifacts on the reconstructed images are plotted for different methods to quantitatively evaluate their reconstruction performance. For the low-energy CT image reconstructed from the projections within $[0, \pi/2)$, the line profile starts from the top right pixel and ends with the bottom left pixel. For the high-energy CT image reconstructed from the projection within $[\pi/2, \pi)$, the line profile starts from the top left pixel and ends with the bottom right pixel. The root mean square error (RMSE), peak-signal-to-noise ratio (PSNR), and structural similarity (SSIM) are also calculated to provide more quantitative comparisons. To further evaluate the performance of different methods, the decomposition results based on the reconstructed DECT images are also generated in this work by using a material decomposition method [51].

## IV. Results

### A. Fusion CT Results

In this work, we first evaluate the benefit of fusion CT in the reconstruction of dual limited-angle problems. Fig. 5(a) shows the fusion CT results of simulated and real data. Figs. 5(b) and 5(c) show the limited-angle low- and high-energy images reconstructed from the collected projections of dual quarter scans via FBP method without enforcing the negative values to zero. Top and bottom row represents the results of simulated and real data, respectively. We can find from Fig. 5 that the results directly reconstructed from the dual limited-angle projections obtain poor quality, and most image information is destroyed by limited-angle artifacts. However, the fusion CT image largely reduces the directional streak artifacts caused by the deficiency of scanning angle and efficiently suppresses the additional black and white artifacts. Evidently, the image edges and inner structures of fusion CT images are clearer than those of the initial reconstruction images. Thus, on the basis of the fusion CT results of simulated and real data, we can conclude that the operation of integrating limited-angle DECT images is beneficial for the limited-angle reconstruction of dual quarter scans.

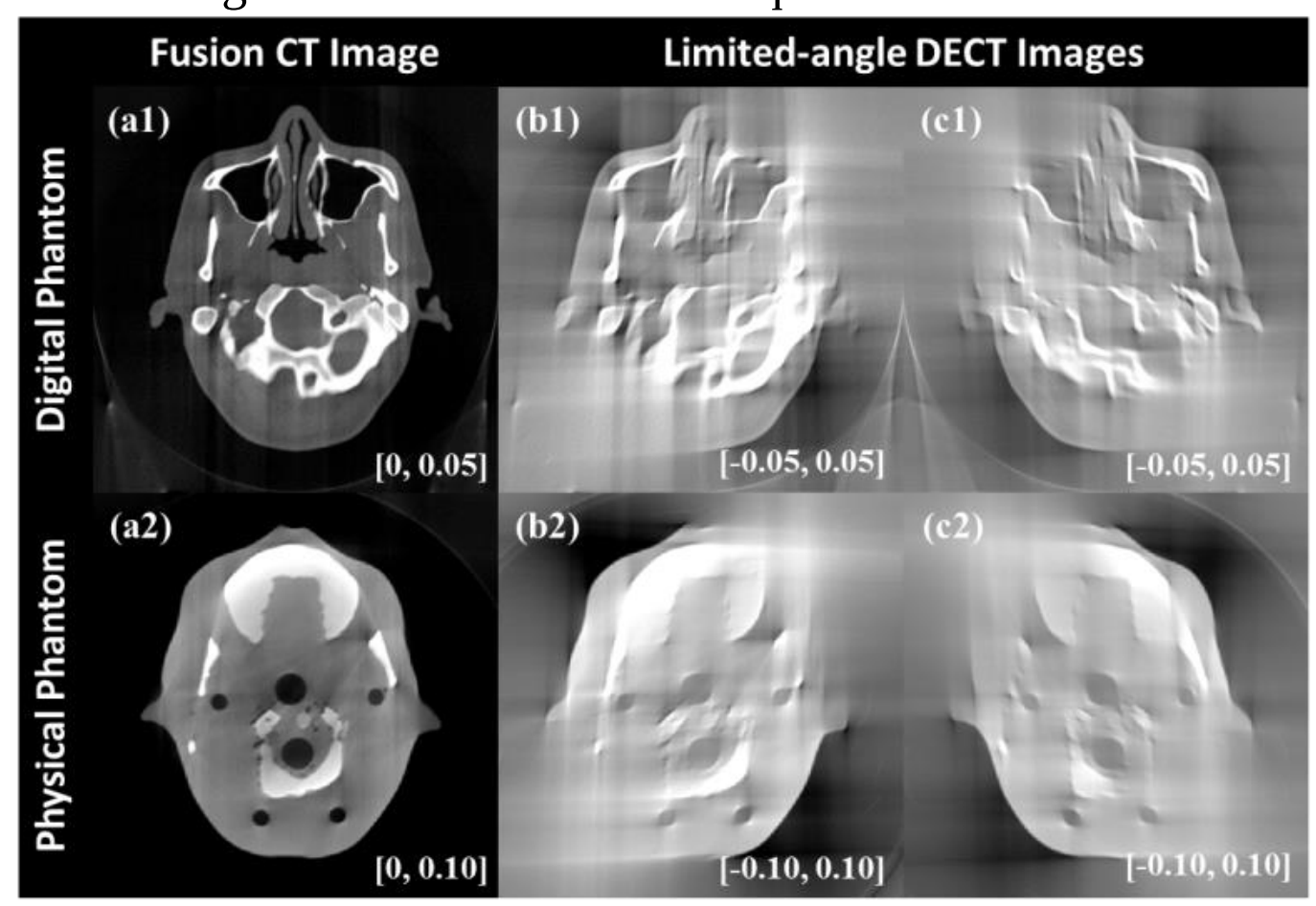

Fig. 5. (a) Fusion CT image. (b) Low- and (c) high-energy results directly reconstructed from the limited-angle projections under dual quarter scans scheme. Top and bottom row represents the results of simulated and real data, respectively. The display windows are denoted in the right bottom of all figures.

### B. Simulated Data Results

Fig. 6 shows the reconstructed results of simulated data at high- and low-energy spectra with different methods. Yellow dashed rectangle denotes the selected paranasal ROI for detailed comparison. Left to right column represents the results of the reference image and the FBP, iterative, learning-based, and proposed methods. We can find that the FBP method fails to reconstruct DECT images from the limited-angle projection data. The regions denoted by the gray arrows in Figs. 6(b1) and 6(b3) are easily influenced by the deficiency of scanning angles. The iterative method reduces the streak artifacts to some extent, but reconstruction images still show obvious degradation in the susceptible regions. The learning-based method improves reconstruction quality and obtains better results than the analytic and iterative methods, but the inner regions obtained via the learning-based method introduce false information and lead to the distortion of image structure, which will influence the judgment of clinical diagnosis. Compared with the former methods, the proposed method efficiently suppresses the limited-angle artifacts and obtains the best reconstruction results. The streak artifacts are invisible, and the image structures are clearly maintained. As shown by the red arrow in Fig. 6, the inner structures of paranasal region are recovered accurately with the proposed method, whereas those obtained by other methods have poor quality. We can also find that although the compared learning-based method can suppress limited-angle artifacts to some extent, it fails to restore the information those have been destroyed by limited-angle artifacts. However, the proposed method recovers the image information with high quality, which validates the benefits of the fusion CT image in this work. Fig. 7 shows the bias maps of the reconstruction results with the reference images. The DECT images of FBP method have the largest differences followed by the iterative and learning-based methods. The proposed method has the smallest image bias among the compared methods.

The line profiles of the reconstruction images are provided to quantitatively compare the accuracy of different methods. In Fig. 8, the line profiles of the FBP and iterative methods depart from the reference lines at low- and high-energy spectra. The learning-based method obtains more accurate lines than the two former conventional methods. The proposed method generates the line profiles closest to the reference results among the compared methods. Especially in the inner region with complex structures, the line profiles of the proposed method are more accurate than those of the compared methods, thereby showing the advantage of the proposed method in the maintenance of detailed structures. TABLE I shows the quantitative comparison results of the reconstructed DECT images for different methods. The FBP method receives the largest RMSEs followed by the iterative and learning-based methods. The proposed method reduces RMSE by two orders of magnitude compared with the FBP method, and by one order of magnitude compared with the iterative and learning-based methods. For the evaluation of PSNR and SSIM, the proposed method also obtains the highest values among the compared methods.

The decomposition results of tissue and bone materials are also generated in Fig. 9 to further evaluate the performance of different methods. Left to right columns denote the decomposition images based on the results of the reference images and the FBP, iterative, learning-based, and proposed methods, correspondingly. Material decomposition is very sensitive to image artifacts and noises. Subtle errors will lead to large biases on the final decomposed results. Due to the interferences of limited-angle artifacts in the reconstruction images, the decomposition results of the FBP and iterative methods have poor quality, and the tissue and bone material images are mixed with each other. Although the learning-based method improves decomposition results to some extent, its results still depart from the reference results. By contrast, the basis material images decomposed from the reconstructed DECT images with the proposed method are with the highest quality and closest to the reference results. In summary, the evaluations of the decomposition results also verify the effectiveness of the proposed method in the dual limited-angle reconstructions under the designed dual quarter scans scheme.

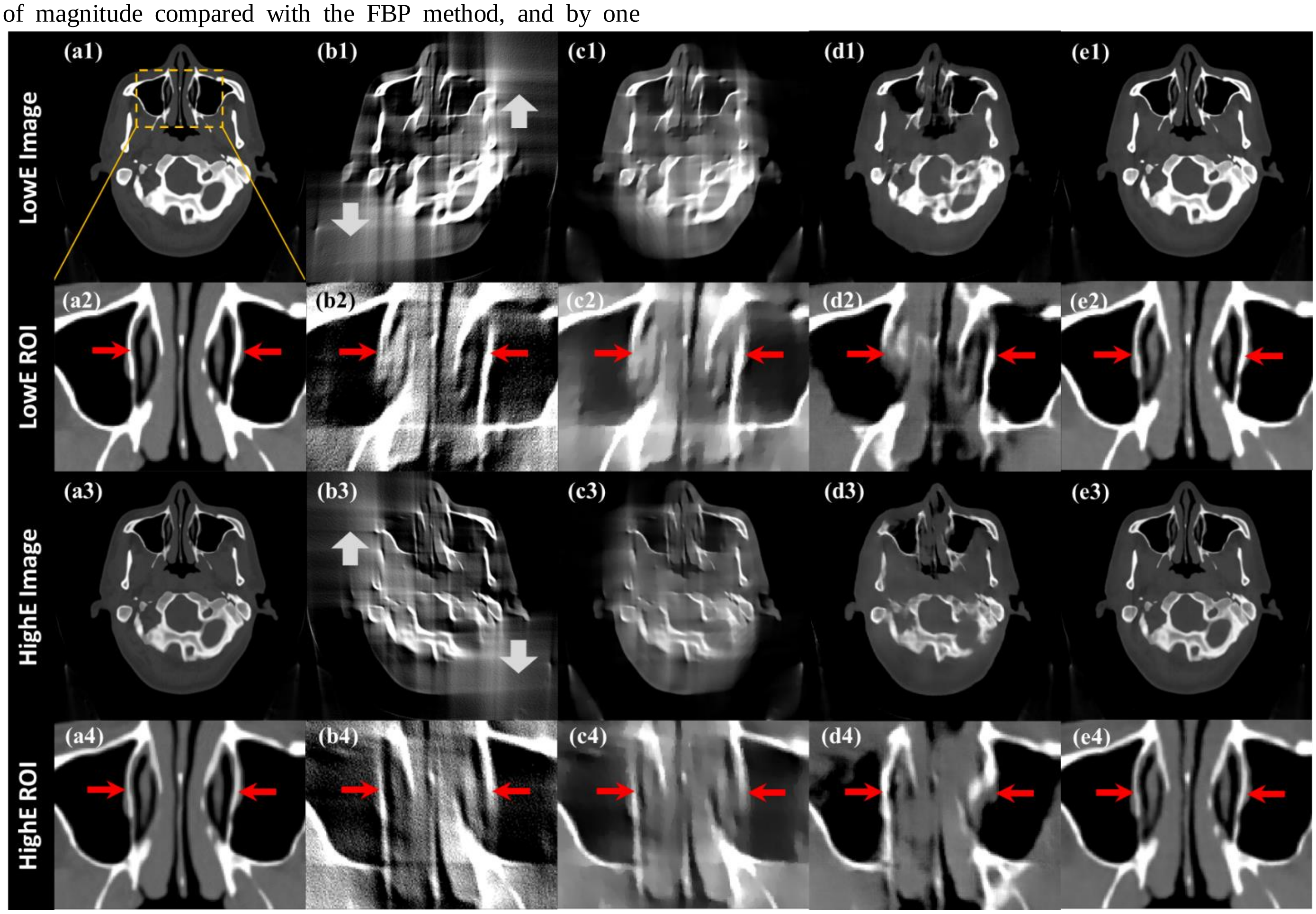

Fig. 6. Reconstruction results of digital phantom from dual quarter scans with different methods. Top to bottom rows represent the reconstructed low-energy CT images, ROIs of low-energy CT images, reconstructed high-energy CT images, and ROIs of high-energy CT images, correspondingly. Left to right columns represent the results of the reference image and the FBP, iterative, learning-based, and proposed methods, correspondingly. The display window of the reconstruction results and the ROIs is [0, 0.05] and [0, 0.03], respectively.

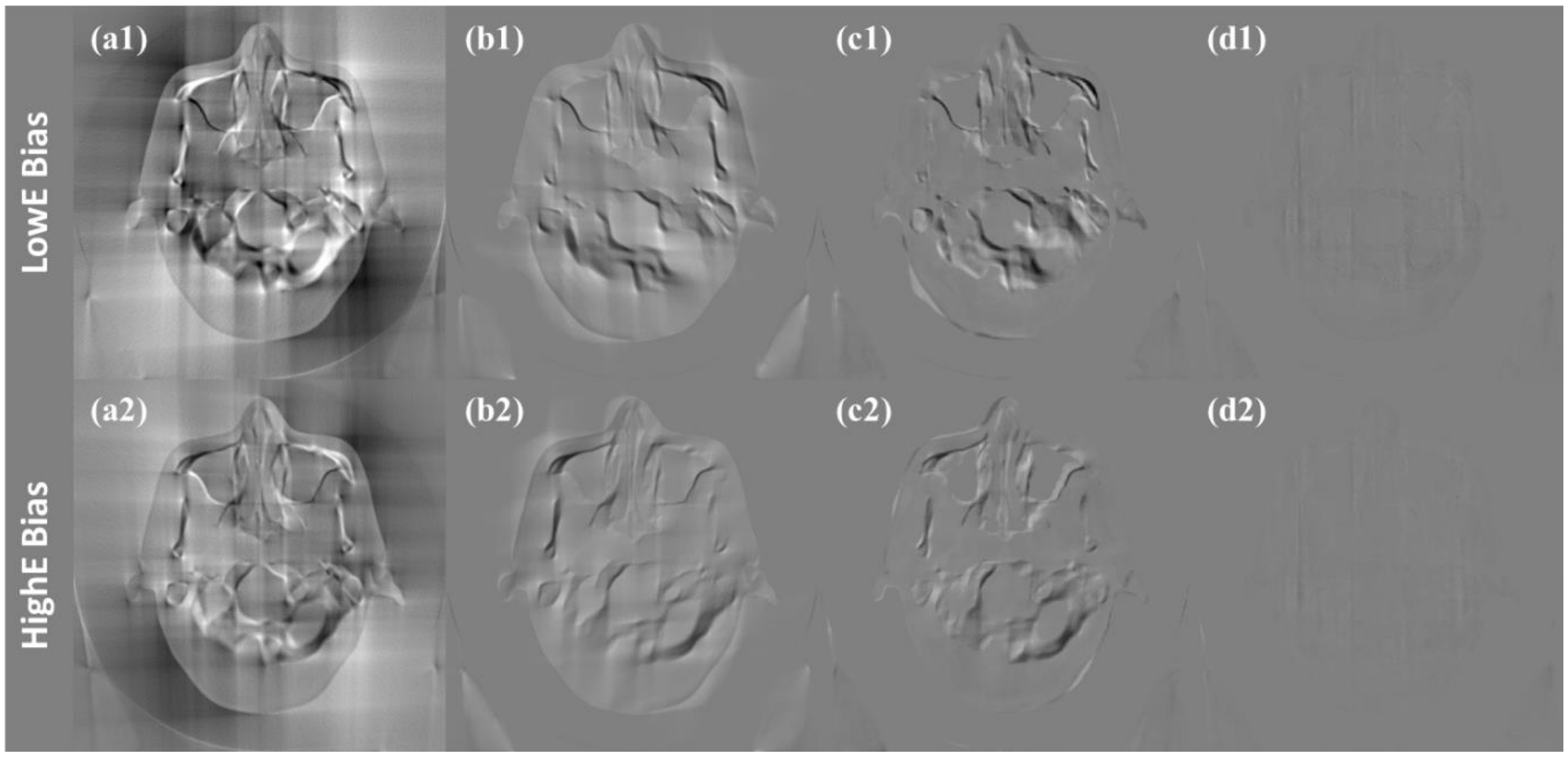

Fig. 7. Bias maps of the reconstructed results of digital phantom via different methods. Top and bottom row represents the bias maps of low- and high-energy CT image, respectively. Left to right columns represent the results of the FBP, iterative, learning-based, and proposed methods, correspondingly. The display windows of all figures are [−0.05, 0.05].

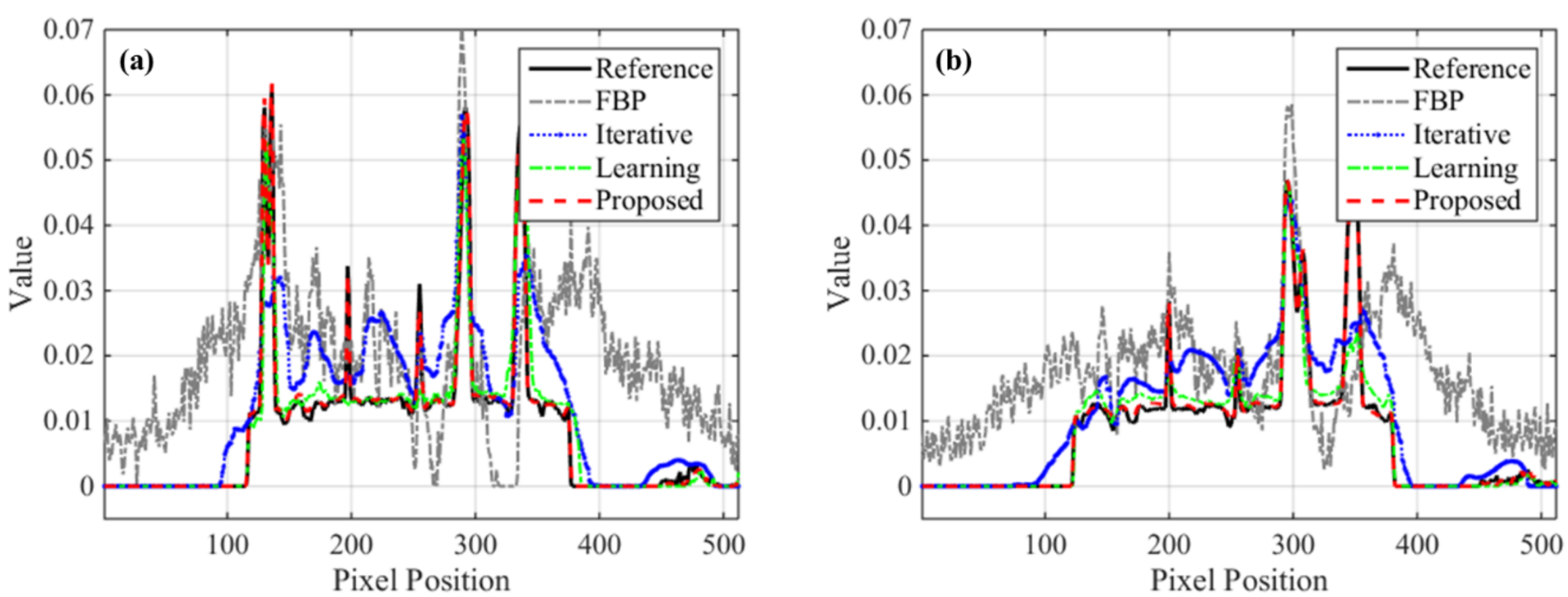

Fig. 8. Line profiles of (a) low- and the (b) high-energy images of digital phantom. Black, gray, blue, green, and red lines represents the results of the reference images and the FBP, iterative, learning-based, and proposed methods, respectively.

TABLE I

RMSEs, PSNRs, AND SSIMs OF THE DIGITAL PHANTOM IMAGES RECONSTRUCTED USING DIFFERENT METHODS.

| Results | Metrics | FBP | Iterative | Learning-based | Proposed |
|---|---|---|---|---|---|
| Low-energy Image | RMSE | 0.0125 | 0.0063 | 0.0044 | $6.4401 \times 10^{-4}$ |
| | PSNR | 38.0700 | 44.0351 | 47.2255 | 63.8222 |
| | SSIM | 0.7725 | 0.9428 | 0.9814 | 0.9993 |
| High-energy Image | RMSE | 0.0096 | 0.0047 | 0.0032 | $5.4107 \times 10^{-4}$ |
| | PSNR | 40.3753 | 46.5590 | 49.9260 | 65.3349 |
| | SSIM | 0.8109 | 0.9557 | 0.9853 | 0.9994 |

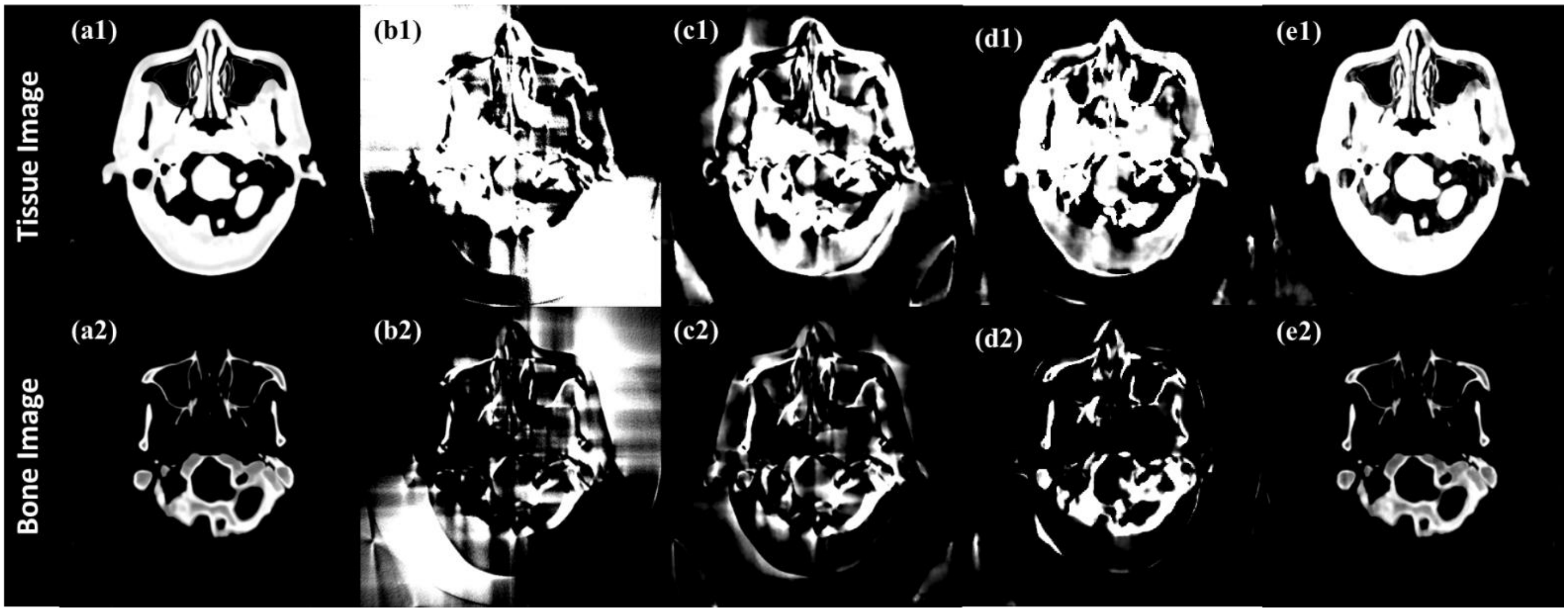

Fig. 9. Decomposition results based on the reconstruction images of digital phantom with different methods. Top and bottom row represents the decomposed tissue and bone materials, respectively. Left to right columns represent the decomposition results based on the results of the reference images and the FBP, iterative, learning-based, and proposed methods, correspondingly. The display windows of all figures are [0.2, 1.0].

### C. *Real Data Results*

Fig. 10 shows the reconstructed DECT images of real data via different methods. Left to right columns represent the results of the reference image and the FBP, iterative, learning-based, and proposed methods, correspondingly. Yellow dashed rectangle represents the magnified ROI of the real data for detailed comparison. The results of the FBP and iterative methods are seriously destroyed by limited-angle artifacts at high- and low-energy spectra. The learning-based method receives better results than the two former conventional methods, but the inner structure of its results are distorted, as shown in Figs. 10(d2) and 10(d4). The proposed method obtains the best reconstruction results at high- and low-energy spectra among the compared methods. It efficiently removes the limited-angle artifacts and maintains the information of the image edges and inner structures for the scanning object. Fig. 11 shows the bias maps of the reconstruction results of different methods. We can intuitively find that the proposed method obtains the smallest bias with the reference images among the compared methods. Although the learning-based method receives smaller biases than the FBP and iterative methods, its bias values along the direction of limited-angle artifacts are still larger than those of the proposed method. The comparison of the learning-based and proposed methods in real data further implies the remarkable role of the fusion CT image in the reconstruction of dual limited-angle problems for the new DECT scanning scheme.

To quantitatively evaluate the accuracy of different methods in real data experiment, Fig. 12 provides the line profiles of reconstruction results along the direction of limited-angle artifacts. The line profiles of the FBP and iterative methods show large differences with those of the reference images at high- and low-energy spectra. The learning-based and proposed methods receive better results than the two former methods. However, when comparing the reconstruction accuracy of detailed information, the proposed method obtains more accurate results than the learning-based method, which are denoted by the black arrows in Fig. 12. TABLE II lists the RMSEs, PSNRs, and SSIMs of the reconstructed results of different methods. Compared with the FBP, iterative, and learning-based methods, the proposed method reduces RMSE by 96.03%, 93.64%, and 74.10%, respectively, on the low-energy CT images and by 96.60%, 94.20%, and 71.63%, respectively, on the high-energy CT images. In terms of the evaluation of PSNR and SSIM, the proposed method also receives the highest values followed by the learning-based, iterative, and FBP methods.

Fig. 13 shows the decomposition results based on the reconstruction images of real data with different methods. The results of the FBP and iterative methods are poor such that the region of the tissue and bone materials cannot be intuitively determined. For the learning-based method, although it receives relatively better reconstruction results in Fig. 10, its decomposition results show a substantial decline. The inner structures and edge information of the decomposition results of the learning-based method are destroyed and many image pixels are incorrectly decomposed on the bone and tissue material images. Compared with the former methods, the proposed method generates better results and maintains clearer image edges and inner structures on the decomposed basis materials. The region of different basis materials can be intuitively determined based on the decomposed results of the proposed method. The comparison of decomposition results further illustrates that the proposed method generates limited-angle DECT images with the highest quality among the compared methods. In summary, consistent with the evaluation conclusion of simulated data, the experimental results of real data also validate the effectiveness of the proposed method in the dual limited-angle reconstruction under dual quarter scans.

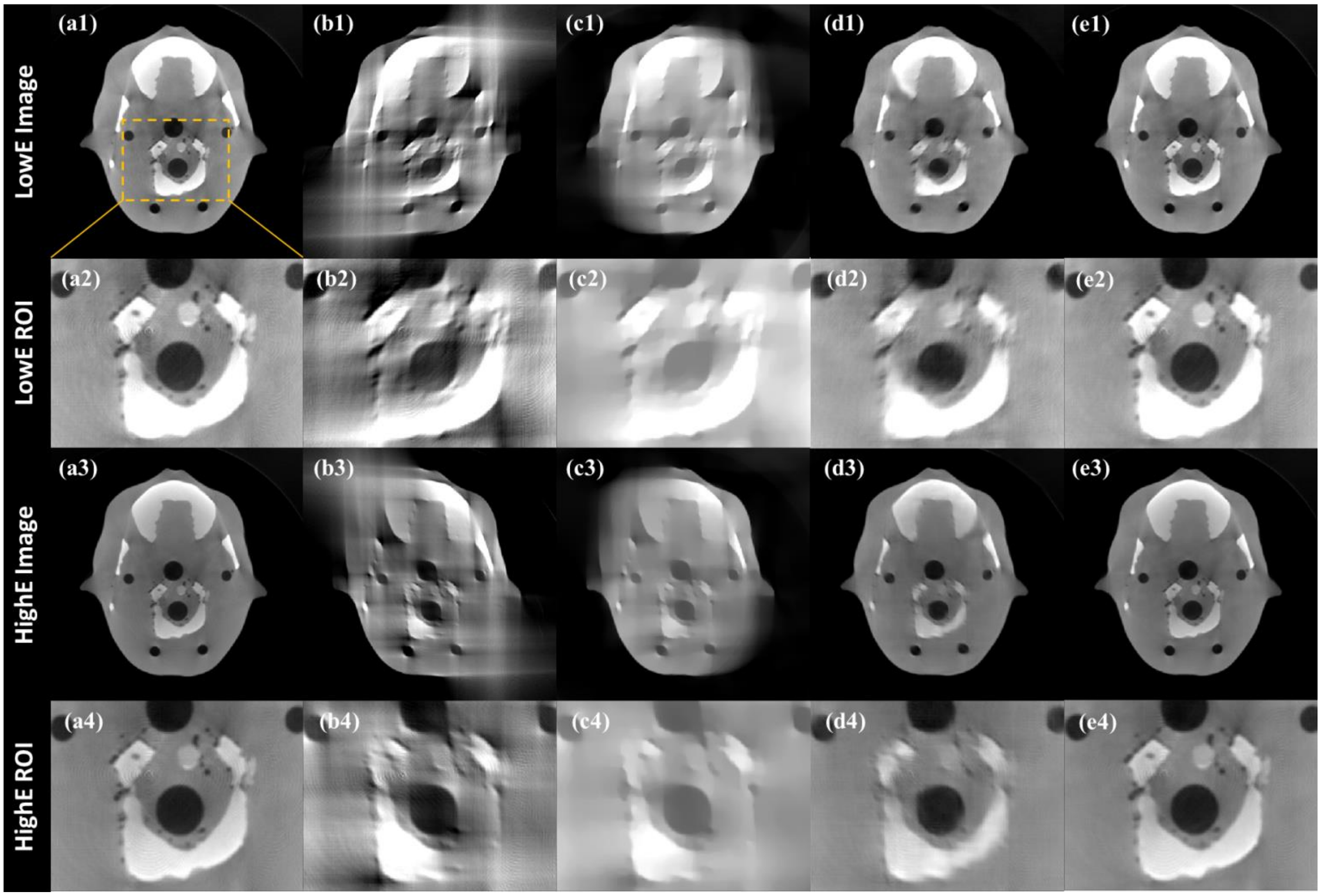

Fig. 10. Reconstruction results of real data from dual quarter scans with different methods. Top to bottom rows represent the reconstructed low-energy CT images, ROIs of low-energy CT images, high-energy CT images, and ROIs of high-energy CT images, correspondingly. Left to right columns represent the results of the reference images and the FBP, iterative, learning-based, and proposed methods, correspondingly. The display window of the reconstruction results and the ROIs is [0, 0.1] and [0, 0.08], respectively.

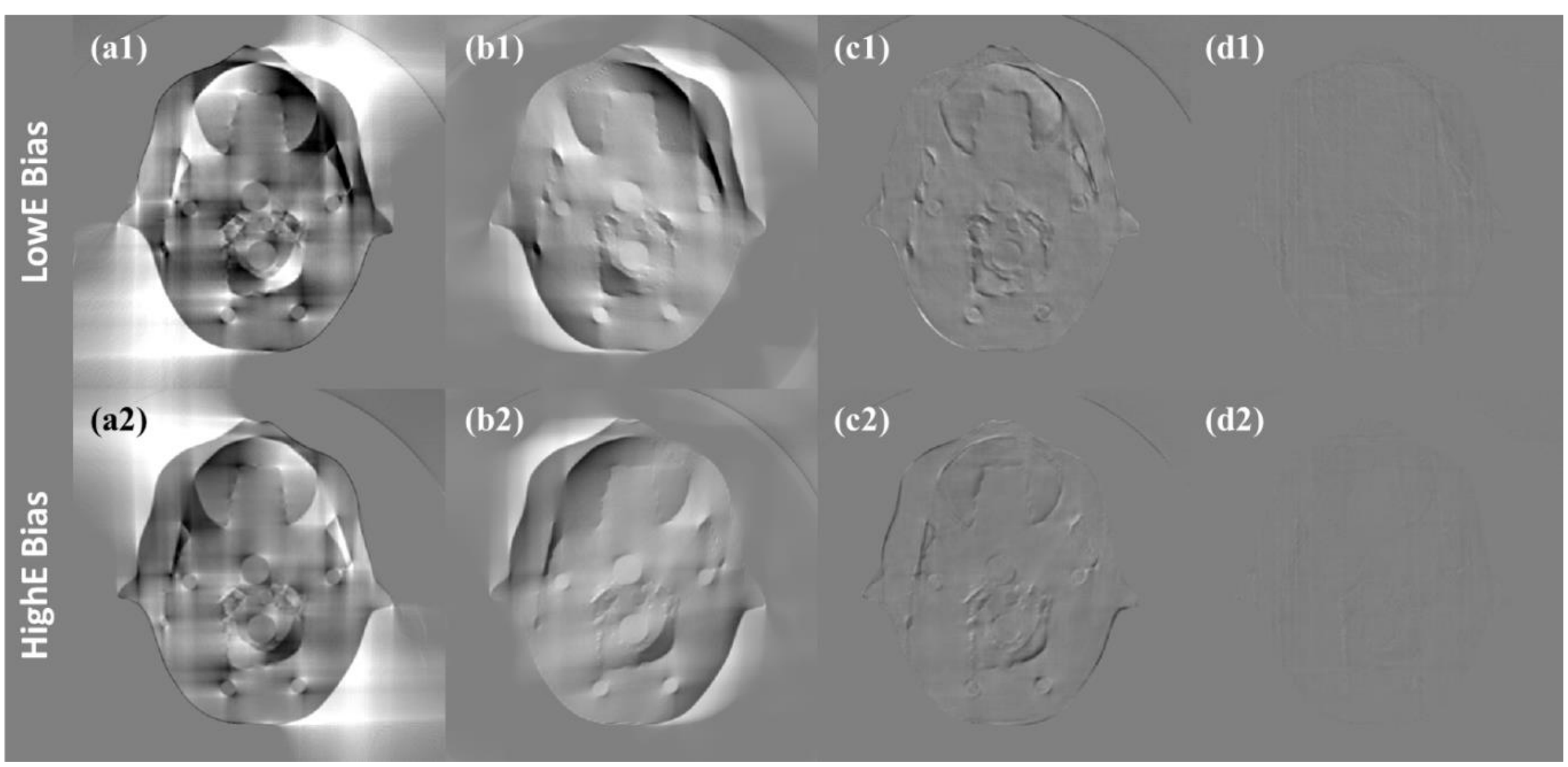

Fig. 11. Bias maps of the reconstructed results of real data via different methods. Top and bottom row represents the bias maps of low- and high-energy CT image, respectively. Left to right columns represent the results of the FBP, iterative, learning-based, and proposed methods, correspondingly. The display windows of all figures are [−0.05, 0.05].

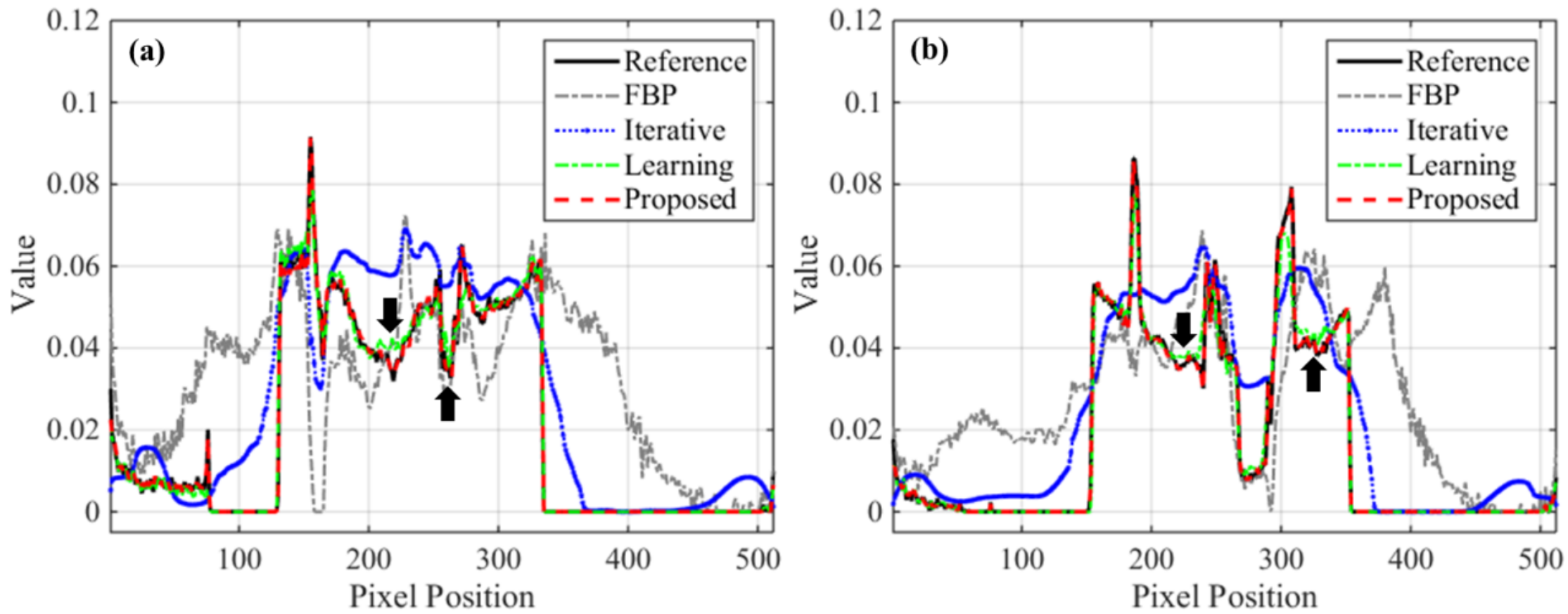

Fig. 12. Line profiles of (a) low- and (b) high-energy images of real data. Black, gray, blue, green, and red lines represents the results of the reference images and the FBP, iterative, learning-based, and proposed methods, respectively.

TABLE II
RMSEs, PSNRs, AND SSIMs OF THE REAL DATA IMAGES RECONSTRUCTED USING DIFFERENT METHODS.

| Results | Metrics | FBP | Iterative | Learning-based | Proposed |
|---|---|---|---|---|---|
| Low-energy Image | RMSE | 0.0184 | 0.0116 | 0.0027 | $6.9921 \times 10^{-4}$ |
| | PSNR | 34.6917 | 38.6921 | 51.5191 | 63.1079 |
| | SSIM | 0.8134 | 0.8569 | 0.9947 | 0.9996 |
| High-energy Image | RMSE | 0.0146 | 0.0084 | 0.0018 | $5.1073 \times 10^{-4}$ |
| | PSNR | 36.7137 | 41.4869 | 54.8424 | 65.8362 |
| | SSIM | 0.8274 | 0.9141 | 0.9974 | 0.9997 |

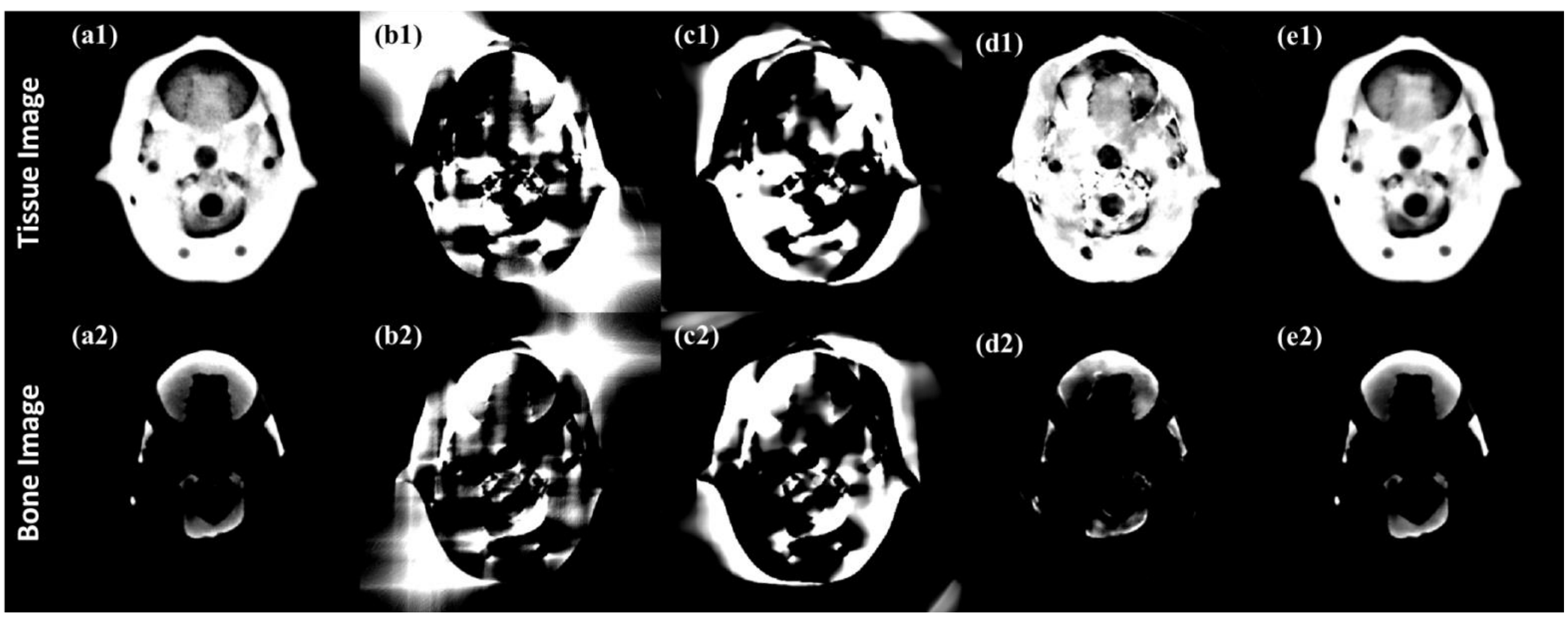

Fig. 13. Decomposition results based on the reconstruction images of real data with different methods. Top and bottom row represents the decomposed tissue and bone materials, respectively. Left to right columns represent the decomposition results based on the results of the reference images and the FBP, iterative, learning-based, and proposed methods, correspondingly. The display window of tissue and bone image is [0.2, 1.0] and [0.4, 1.0], respectively.

## V. DISCUSSION AND CONCLUSION

In this work, we proposed a reconstruction method to enable DECT on the designed dual quarter scans scheme. To solve the dual limited-angle problems caused by the partial-scan, we analyzed the characteristics of image artifacts and found that the limited-angle artifacts of high- and low-energy CT images are complementarily distributed in image domain because the corresponding X-ray lines of dual quarter scans are orthogonal. Inspired by this finding, a fusion CT image was generated by integrating the limited-angle DECT images, which largely reduced the streak artifacts and preserved the image structures and edges well. A novel neural network named as AnNet was further developed in this work to establish the mapping relationship between the fusion CT image and the DECT images. Experimental evaluation on both simulated and real data verified the effectiveness of the proposed method.

The success of the proposed method mainly stems from the generation of fusion CT image inspired by the characteristic analysis of image artifacts. When compared with the same type of learning-based method, the proposed method incorporated with fusion CT image shows evident advantages in the restoration of image information that have already been destroyed by limited-angle artifacts. In previous work, although the characteristics of limited-angle artifacts were already studied, they might have not been fully utilized by researchers. Now we provide a feasible instance by exploring the characteristic of limited-angle artifacts in this work to perform DECT reconstruction from dual quarter scans. In this work, an efficient neural network is developed to yield the DECT images from the fusion CT image. The total running time for one slice of image is less than 0.1 second. Other techniques can also be applied to explore the prior knowledge of fusion CT image as the regularization terms, such as the gradient magnitude map, wavelet basis, and dictionary atom. However, one problem should be considered is that the computation time of the iterative method are usually longer than the neural network, which might restrict its applications in practice.

For the AnNet training, two generation pathways for high- and low-energy images are integrally trained in our work, reducing the training computation and time. The experimental results illustrate that the output results of AnNet are with good quality for the imaging task of dual quarter scans. But we may note that separately training the two generation pathways is also feasible if dealing with complex tasks which need further improvement on the accuracy of the neural network. However, this strategy leads to long training period and large computational memory. For the generation of training dataset, the tube voltages of low- and high-energy are 80 and 140 kVp, respectively, and expanding the training dataset is necessary if you prefer to perform DECT under dual quarter scans at other energies. However, for a specific DECT system, fixed tube voltage and datasets are acceptable because its scanning tube voltage and parameters will not change frequently. For the parameter selection of $\alpha$ in Eq. (1), it is fixed to 0.5 in the experiments of simulated and real data. In fact, $\alpha$ value shows less influence on the final results because the differences of gray value are relatively small in this work for the high- and low-energy CT images. When their differences are large, we may adjust $\alpha$ to obtain a promising fusion CT image. Nevertheless, few parameters need to be adjusted in the proposed method, which enables it more practical in DECT.

In conclusion, this work designed a novel DECT scheme with dual quarter scans, and proposed an efficient method to reconstruct DECT images from dual limited-angle projections by simultaneously utilizing the characteristics of limited-angle artifacts and taking the advantage of neural network in nonlinear mapping. The proposed approach provides a flexible scheme to realize DECT imaging with less requirements of data acquisition and largely reduces the scanning angles and imaging doses. The proposed approach considered as an image domain method can be directly applied to existing imaging configurations with less hardware modification. It is also possible to derive other efficient methods on the basis of the formation of fusion CT image for the potential advanced DECT imaging systems.

## ACKNOWLEDGMENT

This work was supported by the National Natural Science Foundation of China (Grant No. 61601518) and the National Science Foundation for Post-doctoral Scientists of China (Grant No. 2019M663996).

## REFERENCES

[1] W. A. Kalender, W. H. Perman, J. R. Vetter, and E. Klotz, “Evaluation of a prototype dual-energy computed tomographic apparatus. I. Phantom studies,” Med. Phys., vol. 13, no. 3, pp. 334–339, 1986.

[2] T. Johnson, B. Krau, M. Sedlmair, M. Grasruck, H. Bruder, D. Morhard, C. Fink, S. Weckbach, M. Lenhard, B. Schmidt, T. Flohr, M. F. Reiser, and C. R. Becker, “Material differentiation by dual energy CT: initial experience,” European J. Radiology, vol. 17, no. 6, pp. 1510–1517, 2007.

[3] S. Singh and M. Singh, “Explosives detection systems (EDS) for aviation security,” Signal Process., vol. 83, no.1, pp. 31–55, 2003.

[4] Z. Ying, R. Naidu, and C. R. Crawford, “Dual energy computed tomography for explosive detection,” J. X-Ray Sci. Tech., vol. 14, no. 4, pp. 235–256, 2006.

[5] P. Engler and W. Friedman, “Review of dual-energy computed tomography techniques,” Mater. Eval., vol. 48, no. 5, pp. 623–629, 1990.

[6] P. R. Mendonca, P. Lamb, and D. V. Sahani, “A flexible method for multi-material decomposition of dual-energy CT images,” IEEE Trans. Med. Imag., vol. 33, no. 1, pp. 99–116, 2014.

[7] C. H. McCollough, S. Leng, L. Yu, and J. G. Fletcher, “Dual- and multienergy CT: Principles, technical approaches, and clinical applications,” Radiology, vol. 276, no. 3, pp. 637–653, 2015.

[8] Y. Long and J. A. Fessler, “Multi-material decomposition using statistical image reconstruction for spectral CT,” IEEE Trans. Med. Imag., vol. 33, no. 8, pp. 1614–1626, Aug. 2014.

[9] J. Noh, J. A. Fessler, and P. E. Kinahan, “Statistical sinogram restoration in dual-energy CT for PET attenuation correction,” IEEE Trans. Med. Imag., vol. 28, no. 11, pp. 1688–1702, Nov. 2009.

[10] Y. Xue, R. Ruan, X. Hu, Y. Kuang, J. Wang, Y. Long, and T. Niu, “Statistical image-domain multi-material decomposition for dual-energyCT,” Med. Phys., vol. 44, no. 3, pp. 886–901, 2017.

[11] Z. Li, S. Ravishankar, Y. Long, and J. A. Fessler, “DECT-MULTRA: Dual-energy CT image decomposition with learned mixed material models and efficient clustering,” IEEE Trans. Med. Imag., 10.1109/TMI.2019.2946177, Oct. 2019.

[12] M. Daniele, T. B. Daniel, M. Achille, and C. N. Rendon, “State of the art: Dual-energy CT of the abdomen,” Radiology, vol. 271, no. 2, pp. 327–342, 2014.

[13] P. Sukovic and N. H. Clinthorne, “Penalized weighted least-squares image reconstruction for dual energy x-ray transmission tomography,” IEEE Trans. Med. Imaging, vol. 19, no. 11, pp. 1075–1081, 2000.

[14] C. Maaß, M. Baer, M. Kachelrieß, “Image-based dual energy CT using optimized precorrection functions: a practical new approach of material decomposition in image domain,” Med Phys., vol. 36, no.8, pp. 3818–3829, July 2009.

[15] L. Yu, S. Leng and C. H. McCollough, “Dual-energy CT-based monochromatic imaging” Am. J. Roentgenol, vol. 199, no. 5_supplement, pp. S9–S15, 2012.

[16] X. Dong, T. Niu and L. Zhu, “Combined iterative reconstruction and image-domain decomposition for dual energy CT using total-variation regularization,” Med. Phys., vol. 41, no. 5, 051909, 2014.

[17] J. Harms, T. Wang, M. Petrongolo, T. Niu and L. Zhu, “Noise suppression for dual-energy CT via penalized weighted least-square optimization with similarity-based regularization,” Med. Phys., vol. 43, no. 5, pp. 2676–2686, 2016.

[18] M. Petrongolo and L. Zhu, “Noise suppression for dual-energy CT through entropy minimization,” IEEE Trans. Med. Imaging, vol. 34, no. 11, pp. 2286–2297, 2015.

[19] L. Shen and Y. Xing, "Multienergy CT acquisition and reconstruction with a stepped tube potential scan," Med. Phys., vol. 42, no. 1, pp. 282–296, Jan. 2014

[20] T. Wang and L. Zhu, "Dual energy CT with one full scan and a second sparse-view scan using structure preserving iterative reconstruction (SPIR)," Phys. Med. Biol., vol. 61, no. 18, pp. 6684–6706, 2016.

[21] M. Petrongolo and L. Zhu, "Single-Scan Dual-Energy CT Using Primary Modulation," IEEE Trans. Med. Imaging, vol. 37, no. 8, pp. 1799–1808, 2018.

[22] H. Zhang and Y. Xing, "Reconstruction of limited-angle dual-energy CT using mutual learning and cross-estimation (MLCE)," in Proc. SPIE Medical Imaging, 2016, pp. 9783.

[23] B. Chen, Z. Zhang, E. Y. Sidky, D. Xia, and X. Pan, "Image reconstruction and scan configurations enabled by optimization based algorithms in multispectral CT," Phys. Med. Biol., vol. 62, no. 22, pp. 8763–8793, 2017.

[24] W. Zhang, L. Wang, L. Li, T. Niu, Z. Li, N. Liang, Y. Xue, B. Yan, and G. Hu, "Reconstruction method for DECT with one half-scan plus a second limited-angle scan using prior knowledge of complementary support set (Pri-CSS)" Med. Phys., vol. 65, no. 2, 025005, 2020.

[25] J. Wang, L. Zeng, C. Wang, and Y. Guo, "ADMM-based deep reconstruction for limited-angle CT," Phys. Med. Biol., vol. 64, no. 11, 115011, 2019.

[26] E. Y. Sidky and X. Pan, "Image reconstruction in circular cone-beam computed tomography by constrained, total-variation minimization," Phys. Med. Biol., vol. 53, no. 17, 4777, 2008.

[27] Z. Chen, X. Jin, L. Li, and G. Wang, "A limited-angle CT reconstruction method based on anisotropic TV minimization," Phys. Med. Biol., vol. 58, no. 7, pp. 2119–2141, 2013.

[28] Y. Quan, H. Ji, and Z. Shen, "Data-driven multi-scale non-local wavelet frame construction and image recovery," J. Sci. Comput., vol. 63, no. 2, pp. 307–329, 2014.

[29] A. Cai, L. Li, Z. Zheng, H. Zhang, L. Wang, G. Hu, and B. Yan, "Block matching sparsity regularization-based image reconstruction for incomplete projection data in computed tomography," Phys. Med. Biol., vol. 63, no. 3, 035045, 2018.

[30] D. Wu and L. Zeng, "Limited-angle reverse helical cone-beam CT for pipeline with low rank decomposition," Opt. Commun., vol. 328, no. 10, pp. 109–115, 2014.

[31] M. Cao and Y. Xing, "Limited angle reconstruction with two dictionaries," in Proc. IEEE Nuclear Science Symposium and Medical Imaging Conference (NSS/MIC'13), Oct. 2013, Seoul, Republic of Korea, pp. 1–4.

[32] E. T. Quinto, "Tomographic reconstructions from incomplete data numerical inversion of the exterior Radon transform," Inverse Probl., vol. 4, no. 3, pp. 867–876, 1988.

[33] J. Frikel and E. T. Quinto, "Characterization and reduction of artifacts in limited angle tomography," Inverse Probl., vol. 29, no. 12, 125007, 2013.

[34] L. V. Nguyen, "How strong are streak artifacts in limited angle computed tomography?" Inverse Probl., vol. 31, no. 5, 055003, 2015.

[35] G. Wang, "A perspective on deep imaging," IEEE Access, vol. 4, pp. 8914–8924, Nov. 2016.

[36] E. Kang, W. Chang, J. Yoo, and J. C. Ye, "Deep Convolutional Framelet Denosing for Low-Dose CT via Wavelet Residual Network," IEEE Trans. Med. Imag., vol. 37, no. 6, pp. 1358–1369, Apr. 2018.

[37] R. Anirudh, H. Kim, J. J. Thiagarajan, K. A. Mohan, K. M. Champley, and T. Bremer, "Lose the views: Limited angle CT reconstruction via implicit sinogram completion," in Proc. IEEE Conference on Computer Vision and Pattern Recognition (CVPR), June 2018, Salt Lake City, Utah. DOI: 10.1109/CVPR.2018.00664.

[38] H. Zhang, L. Li, K. Qiao, L. Wang, B. Yan, L. Li, and G. Hu, "Image prediction for limited-angle tomography via deep learning with convolutional neural network," arXiv: 1607.08707, 2016.

[39] J. Gu and J. Ye, "Multi-scale wavelet domain residual learning for limited-angle CT reconstruction," arXiv:1703.01382, 2017.

[40] J. Zhao, Z. Chen, L. Zhang, X. Jin, "Unsupervised learnable sinogram inpainting network (SIN) for limited angle CT reconstruction," arXiv:1811.03911, 2018.

[41] H. Turbell, "Cone-beam reconstruction using filtered backprojection," Ph. D. dissertation, Linköping Univ., Linköping, Sweden, 2001.

[42] L. Borg, J. Frikel, J. S. Jorgensen, and E. T. Quinto, "Full characterization of reconstruction artifacts from arbitrary incomplete X-ray CT data," arXiv: 1707.03055v6, 2018.

[43] O. Ronneberger, P. Fischer, and T. Brox, "U-Net: Convolutional Networks for Biomedical Image Segmentation," in Proc. The 18th Medical Image Computing and Computer-Assisted Intervention, Oct. 2015, Munich, Germany, pp, 234-241.

[44] K. He, X. Zhang, S. Ren, and J. Sun, "Delving deep into rectifiers: surpassing human-level performance on ImageNet classification," in Proc. IEEE International Conference on Computer Vision (ICCV), Dec. 2015, Santiago, Chile, pp. 1026–1034.

[45] D. Kingma and J. Ba, "Adam: A method for stochastic optimization," arXiv: 1412.6980v8, 2015.

[46] B. Chen, Z. Zhang, D. Xia, and E. Y. Sidky, "Algorithm-enabled partial-angular-scan configurations for dual-energy CT," Med. phys., vol. 45, no. 5, pp. 1857–1870, 2018.

[47] Y. Li, K. Li, C. Zhang, J. Montoya, and G. Chen, "Learning to reconstruct computed tomography (CT) images directly from sinogram data under a variety of data acquisition conditions," IEEE Trans. Med. Imaging, vol. 38, no. 10, pp. 2469–2481, 2019.

[48] Q. D. Man, E. Haneda, B. Claus, P. Fitzgerald, B. D. Man, G. Qian, H. Shan, J. Min, M. Sabuncu, G. Wang, "A two-dimensional feasibility study of deep learning-based feature detection and characterization directly from CT sinograms," Med. Phys., vol. 46, no. 12, pp. e790–e800, Dec. 2019.

[49] Q. Xie, M. Zhou, Q. Zhao, D. Meng, W. Zuo, and Z. Xu, "Multispectral and hyperspectral image fusion by MS/HS fusion net," in Proc. IEEE Conference on Computer Vision and Pattern Recognition (CVPR), June 2019, Long Beach, California, pp. 1585–1594.

[50] H. Yu and G. Wang, "Compressed sensing based interior tomography," Phys. Med. Biol., vol. 54, no. 9, pp. 2791–2805, 2009.

[51] T. Niu, X. Dong, M. Petrongolo, and L. Zhu, "Iterative image-domain decomposition for dual-energy CT," Med. Phys., vol. 41, no. 4, 041901, 2014.